\documentclass[12pt]{article}
\newcommand{\be}{\begin{equation}}
\newcommand{\ee}{\end{equation}}
\newcommand{\ben}{\begin{equation*}}
\newcommand{\een}{\end{equation*}}
\newcommand{\ar}{\begin{array}}
\newcommand{\arn}{\end{array}}

\newcommand{\x}{\vec{r}}
\newcommand{\xs}{\vec{r}^{\;2}}

\def\pnot{\mbox{${\not{\hbox{\kern-3.0pt$p$}}}$}}
\def\qnot{\mbox{${\not{\hbox{\kern-2.0pt$q$}}}$}}
\def\enot{\mbox{${\not{\hbox{\kern-2.0pt$e$}}}$}}
\def\knot{\mbox{${\not{\hbox{\kern-2.0pt$k$}}}$}}

\def\fun#1#2{\lower3.6pt\vbox{\baselineskip0pt\lineskip.9pt\ialign
{$\mathsurround=0pt#1\hfil##\hfil$\crcr#2\crcr\sim\crcr}}}

\begin{document}

\begin{titlepage}
\hfill Budker INP 2012-19
\vspace{0.7cm}

\begin{center}
{\bf Difference between standard and quasi-conformal BFKL kernels~$^{~\ast}$}
\end{center}

\vskip 0.5cm

\centerline{V.S.~Fadin$^{a\,\dag}$, R.~Fiore$^{b\,\ddag}$,
A.~Papa$^{b\,\dag\dag}$}

\vskip .6cm

\centerline{\sl $^{a}$ Budker Institute of Nuclear Physics, 630090 Novosibirsk,
Russia}
\centerline{\sl Novosibirsk State University, 630090 Novosibirsk, Russia}
\centerline{\sl $^{b}$ Dipartimento di Fisica, Universit\`a della Calabria,}
\centerline{\sl and Istituto Nazionale di Fisica Nucleare, Gruppo collegato di
Cosenza,}
\centerline{\sl Arcavacata di Rende, I-87036 Cosenza, Italy}

\vskip 2cm

\begin{abstract}
As it was recently shown, the colour singlet BFKL kernel, taken in M\"{o}bius
representation in the space of impact parameters, can be written in
quasi-conformal shape, which is unbelievably simple compared with the
conventional form of the BFKL kernel in momentum space. It was also proved
that the total kernel is completely defined by its M\"{o}bius representation.
In this paper we calculated the difference between standard and
quasi-conformal BFKL kernels in momentum space and discovered that it
is rather simple. Therefore we come to the conclusion that the
simplicity of the quasi-conformal kernel is caused mainly by using the impact
parameter space.

\end{abstract}


\vfill \hrule \vskip.3cm \noindent $^{\ast}${\it Work supported
in part by the Ministry of Education and Science of Russian Federation, grant  14.740.11.0082 of Federal Program "Personnel of Innovational Russia", in part by  RFBR grant 10-02-01238,  and in part by
Ministero Italiano dell'Istruzione, dell'Universit\`a e della
Ricerca.} \vfill $
\begin{array}{ll} ^{\dag}\mbox{{\it e-mail address:}} &
\mbox{fadin@inp.nsk.su}\\
^{\ddag}\mbox{{\it e-mail address:}} &
\mbox{roberto.fiore@cs.infn.it}\\
^{\dag\dag}\mbox{{\it e-mail address:}} &
\mbox{alessandro.papa@cs.infn.it}\\
\end{array}
$

\end{titlepage}

\vfill \eject

\section{Introduction}

The BFKL (Balitsky-Fadin-Kuraev-Lipatov) approach~\cite{BFKL} was formulated
in the momentum space. In this space the kernel of the BFKL equation was
calculated in the next-to-leading order (NLO) long ago, at first for the
forward scattering (i.e. for $t=0$ and colour singlet in the
$t$-channel)~\cite{Fadin:1998py} and then for any fixed (not growing with
energy) squared  momentum transfer $t$ and any possible two-gluon
colour state in the $t$-channel~\cite{FF05}. Unfortunately, the NLO kernel is
rather complicated. In particular,  the colour singlet kernel for
$t\neq 0$ is found in the NLO in the form of an intricate two-dimensional
integral.

In the most interesting for phenomenological applications case of colourless
particle scattering, the leading-order (LO) BFKL kernel has a remarkable
property~\cite{Lipatov:1985uk}: it can be taken in the M\"{o}bius
representation (i.e. in the space of functions vanishing at coinciding
transverse coordinates of Reggeons), where it turns out to be invariant
in regard to conformal transformations of these coordinates. Moreover, in the
coordinate space the M\"{o}bius representation (we will call it ``M\"{o}bius
form'') of the LO BFKL kernel coincides~\cite{Fadin:2006ha} with the kernel of
the colour dipole model~\cite{dipole}.

In the NLO the conformal invariance is violated in QCD by the running coupling.
One could hope that the M\"{o}bius form of the colour singlet NLO kernel is
quasi-conformal, i.e. conformal invariance is violated only by terms
proportional to the $\beta$-function. However, the direct transformation of
the colour singlet kernel found in Ref.~\cite{FF05} from momentum to
coordinate space with the restriction of M\"{o}bius representation gives a
kernel which is not quasi-conformal~\cite{Fadin:2007ee,Fadin:2007de,
Fadin:2007xy}.
But in the NLO kernel there is an
ambiguity~\cite{Fadin:2006ha,Kovchegov:2006wf}, analogous to the well known
ambiguity of the NLO anomalous dimensions, because it is possible to
redistribute radiative corrections between the kernel and the impact factors.
The ambiguity, discussed in details in Ref.~\cite{Fadin:2009za}, permits to
make transformations
\begin{equation}
\hat{\mathcal{K}}\rightarrow \hat{\mathcal{K}}
-\alpha_s[\hat{\mathcal{K}}^{(B)},\hat{U}]~\label{kernel transformation}
\end{equation}
conserving the LO kernel $\hat{\mathcal{K}}^{(B)}$ (which is fixed in our case
by the requirement of conformal invariance of its M\"{o}bius form) and
changing the NLO part of the kernel. Note that this transformation must
conserve the gauge invariance properties of the kernel, so that the operator
$\hat{U}$ must have in this respect the same properties as
$\hat{\mathcal{K}}^{(B)}$.

The NLO kernel calculated in Ref.~\cite{FF05} is defined according to the
prescriptions given in Ref.~\cite{Fadin:1998fv}. We will call it the
``standard kernel''. Recently it was shown~\cite{Fadin:2009gh} that there
exist an operator $\hat{U}$ such that the 
transformation~(\ref{kernel transformation}) applied to this standard kernel
gives a kernel with quasi-conformal M\"{o}bius form, which agrees with the 
form obtained in Ref.~\cite{Balitsky:2009xg} in the colour dipole approach.  
It turns out that this form is quite simple. It is unbelievably simple in 
comparison with the form of the standard kernel~\cite{FF05}. Evidently, the 
question arose about the relation between these two forms.

This question is not trivial not only because the M\"{o}bius form is defined
in the coordinate space, whereas the standard kernel was calculated 
in the momentum space.
Remind that the M\"{o}bius representation is defined on a special class of
functions. Therefore at the first sight it seemed impossible to reconstruct
the complete operator from its M\"{o}bius form. However, due to the gauge
invariance of the BFKL kernel, it is not so. It was shown~\cite{Fadin:2011jg}
for any gauge invariant two-particle operator that it is possible to restore
the complete operator from its M\"{o}bius form and the restoration is unique
up to terms which do not contribute to the operator matrix elements, because
of symmetry and gauge invariance of the wave functions.

Therefore, it is in principle possible to restore the complete BFKL kernel
from its quasi-conformal M\"{o}bius form. Since this form is quite simple,
one can hope for simplicity of the complete kernel in the momentum space too.
Evidently this kernel differs from the standard kernel found in
Ref.~\cite{FF05}, but is connected with the last one by the
transformation~(\ref{kernel transformation}). However, the direct restoration
is not easy. It includes the Fourier transformation of the M\"{o}bius form
from coordinate to momentum space and, although this form is very compact,
the transformation is intricate since it contains complicated integrals.

Instead, one can try to find the difference between the standard kernel and the
one restored from the quasi-conformal M\"{o}bius form. Our paper is devoted to
the solution of this problem. The difference under investigation is given by
the second term in the transformation~(\ref{kernel transformation}).  
For the operator $\hat{U}$, both M\"{o}bius form and complete representation 
in the momentum space are known now~\cite{Fadin:2011jg}.  The same is true for
$\hat{\mathcal{K}}^{(B)}$. We are looking for the difference in the momentum
space. It can be found using for the calculation of the commutator
in the transformation~(\ref{kernel transformation}) both $\hat{U}$ and
$\hat{\mathcal{K}}^{(B)}$ in this space. Alternatively, it is possible to
calculate the commutator in the coordinate M\"{o}bius space and then to
restore its complete form in the momentum space using the method developed in
Ref.~\cite{Fadin:2011jg}. We use both these ways, on one side for
cross-checking the obtained result, on the other for a demonstration of the
efficiency of the method of restoration of complete operators from their
M\"{o}bius forms, developed in Ref.~\cite{Fadin:2011jg}.

The paper is organized as follows. In the next Section we calculate the
commutator in the transformation~(\ref{kernel transformation}) directly 
in the momentum space. In Section~3 this commutator is calculated firstly in 
the coordinate M\"{o}bius space and then the obtained result is used for 
restoration of the complete form of the commutator in the momentum space. The 
last Section contains our conclusions. The integrals used in the calculations 
are presented in the Appendix.

\section{Direct calculation of the difference in momentum space}
\label{sec:direct}

We adopt the notation used in Ref.~\cite{Fadin:2011jg} and put the space-time
dimension $D$ equal to 4, so that states $|\vec{q}\rangle$ with definite
two-dimensional transverse Reggeon momentum $\vec{q}$ and states
$|\vec {r}\rangle$ with definite Reggeon impact parameter $\vec{r}$ are
normalized as follows:
\begin{equation}
\langle\vec{q}|\vec{q}^{\;\prime}\rangle=\delta(\vec{q}-\vec{q}^{\;\prime
})\;,\;\;\;\;\;\langle\vec{r}|\vec{r}^{\;\prime}\rangle=\delta(\vec{r}-\vec
{r}^{\;\prime})\;,\;\;\;\;\;
\langle\vec{r}|\vec{q}\rangle=\frac{e^{i\vec{q}\,\vec
{r}}}{2\pi}\;.\label{normalization}%
\end{equation}
As it was shown in Ref.~\cite{Fadin:2009gh}, the quasi-conformal
kernel $\hat{\mathcal{K}}^{QC}$ can be obtained from the kernel
calculated in Ref.~\cite{FF05} by the
transformation~(\ref{kernel transformation}), namely,
\begin{equation}
\hat{\mathcal{K}}^{QC} = \hat{\mathcal{K}}-\alpha_s[\hat{\mathcal{K}}^{(B)},
\hat{U}]~.  \label{QC kernel}
\end{equation}
It is worthwhile to note here that the kernel $\hat{\mathcal{K}}$ is defined
in such a way that in the LO its M\"{o}bius form is conformal invariant.
Therefore one has (see Ref.~\cite{Fadin:2011jg} for details)
\begin{equation}
\langle\vec{q}_{1},\vec{q}_{2}|\hat{\mathcal{K}}|\vec{q}_{1}^{\;\prime},\vec
{q}_{2}^{\;\prime}\rangle=\delta(\vec{q}_{1}+\vec{q}_{2}-\vec{q}_{1}%
^{\;\prime}-\vec{q}_{2}^{\;\prime})\frac{1}{\vec{q}_{1}^{\,\,2}\vec
{q}_{2}^{\,\,2}}{K}(\vec{q}_{1},\vec{q}_{1}^{\;\prime};\vec{q})%
\;,\label{relation  between kernels}%
\end{equation}
where $\vec{q}=\vec{q}_{1}+\vec{q}_{2}=\vec{q}_{1}^{\;\prime}+\vec
{q}_{2}^{\;\prime}$ and ${K}(\vec{q}_{1},\vec{q}%
_{1}^{\;\prime};\vec{q})$ is the symmetric kernel
\begin{equation}
{K}(\vec{q}_{1},\vec{q}_{1}^{\;\prime};\vec{q}) ={K}(\vec{q}_{1}^{\;\prime},
\vec{q}_{1};\vec{q})\;, \label{symmetry of the kernel}%
\end{equation}
defined in Ref.~\cite{Fadin:1998fv} and calculated in Ref.~\cite{FF05}.
Its real part ${K}_r$ satisfies the gauge invariance conditions
\begin{equation}
{K}_r(\vec{0},\vec{q}_{1}^{\;\prime};\vec{q}) ={K}_r(\vec{q}_{1},
\vec{0};\vec{q})
={K}_r(\vec{q},\vec{q}_{1}^{\;\prime};\vec{q}) ={K}_r(\vec{q}_{1},
\vec{q};\vec{q})\;. \label{gauge invariance of the kernel}%
\end{equation}
Our goal is to find in the momentum space an explicit form for the
commutator in Eq.~(\ref{QC kernel}). In this Section it is done using the known
expressions in this space for the LO kernel $\hat{\mathcal{K}}^{(B)}$ and the
operator $\hat{U}$.

The kernel $\hat{\mathcal{K}}^{(B)}$ can be presented as follows
\begin{equation}
\langle\vec{q}_{1},\vec{q}_{2}|{\hat{\mathcal{K}}}^{(B)}|\vec{q}%
_{1}^{\,\,\prime},\vec{q}_{2}^{\,\,\prime}\rangle=\delta(\vec{q}_{11^{\prime}%
}+\vec{q}_{22^{\prime}})\frac{\alpha_{s}N_{c}}{2\pi^{2}}\left[R(\vec{q}_{1},
\vec{q}_{2}; \vec{k})-\delta(\vec{k})\int d\vec{l} \; V(\vec{q}_{1},
\vec{q}_{2};  \vec{l}) \right]\;, \label{K B  momentum}%
\end{equation}
where $\vec{k}=\vec{q}_{11^{\prime}}=-\vec{q}_{22^{\prime}}$
(here and below $\vec{a}_{ij^{\prime}}=\vec{a}_{i}-\vec{a}^{\;\prime}_{j},\;
\vec{a}_{ij}=\vec{a}_{i}-\vec{a}_{j}, \;\vec{a}_{i^{\prime}j^{\prime}}
=\vec{a}^{\;\prime}_{i}-\vec{a}^{\;\prime}_{j}$),
\begin{equation}
R(\vec{q}_{1},\vec{q}_{2};  \vec{k})= \frac{2}%
{\vec{k}^{\,2}}-2\frac{\vec{q}_{1}\vec{k}}{\vec{k}^{\,2}\vec{q}%
_{1}^{\,\,2}}+2\frac{\vec{q}_{2}\vec{k}}{\vec{k}^{\,2}\vec{q}_{2}^{\,\,2}%
} -2\frac{\vec{q}_{1}\vec{q}_{2}}{\vec{q}_{1}^{\,\,2}\vec{q}_{2}%
^{\,\,2}}\label{R  momentum}
\end{equation}
and
\begin{equation}
V(\vec{q}_{1},\vec{q}_{2};  \vec{l})=  \frac{2}{\vec{l}^{\,\,2}}%
-\frac{\vec{l}(\vec{l}-\vec{q}_{1})}{\vec{l}^{\,\,2}(\vec{l}-\vec{q}_{1})^{2}%
}-\frac{\vec{l}(\vec{l}-\vec{q}_{2})}{\vec{l}^{\,\,2}(\vec{l}-\vec{q}%
_{2})^{\,2}}\;. \label{V  momentum}%
\end{equation}
Note that the term ${2}/{\vec{l}^{\,\,2}}$ in $V(\vec{q}_{1},\vec{q}_{2};
\vec{l})$ leads to the divergence of the integral over $d\vec{l}$ in the
second term of Eq.~(\ref{K B  momentum}), which represents the virtual part of 
the kernel. It is a well known infrared divergence which cancels with the
divergence coming from the term ${2}/{\vec{k}^{\,2}}$ in the part with
$R(\vec{q}_{1},\vec{q}_{2};  \vec{k})$ in Eq.~(\ref{K B  momentum}) (real 
part), when ${\hat{\mathcal{K}}}^{(B)}$ acts on some state. In the commutator
$[\hat{\mathcal{K}}^{(B)},\hat{U}]$ there are no problems with these
divergences at all, because they cancel separately in the virtual and real
parts.

The gauge invariance properties for $R$  look as follows:
\[
R(\vec{q}_{1},\vec{q}_{2};  \vec{q}_1)=R(\vec{q}_{1},\vec{q}_{2};  -\vec{q}_2)
=0, \;\;
\]
\begin{equation}
( \vec{q}_{1}^{\;2}\vec{q}_{2}^{\;2} R(\vec{q}_{1},\vec{q}_{2};
\vec{k}))|_{\vec{q}_{1}=0}=(\vec{q}_{1}^{\;2}\vec{q}_{2}^{\;2}
R(\vec{q}_{1},\vec{q}_{2};  \vec{k}))|_{\vec{q}_{2}=0}=0~.
\label{g i for R}
\end{equation}

An explicit form of the operator ${\hat{U}}$ in the momentum space was found
in Ref.~\cite{Fadin:2011jg}. Omitting terms which do not contribute to the
commutator in Eq.~(\ref{QC kernel}), we have
\[
\langle\vec{q}_{1},\vec{q}_{2}|\alpha_s{\hat{U}}|\vec{q}_{1}^{\,\,\prime}%
,\vec{q}_{2}^{\,\,\prime}\rangle=\delta(\vec{q}_{11^{\prime}}+\vec
{q}_{22^{\prime}})\frac{\alpha_s N_c}{4\pi^{2}}R_u(\vec{q}_1, \vec{q}_2;
\vec{k})
\]%
\begin{equation}
-\frac{\alpha_s\beta_0}{8\pi} \ln\left(\vec{q}_{1}^{\;2}
\vec{q}_{2}^{\;2}\right)\delta(\vec{q}_{11^{\prime}})\delta(\vec
{q}_{22^{\prime}})~,
\label{U momentum}
\end{equation}
where $\beta_0$ is the first coefficient of the Gell-Mann--Low function,
\begin{equation}
\beta_0 =\frac{11}{3}N_c -\frac{2}{3}n_f
\label{beta 0}
\end{equation}
and
\[
R_u(\vec{q}_1, \vec{q}_2; \vec{k})=
 \frac{1}{\vec{q}_1^{\,\,2}}\ln\left(\frac{\vec{q}_{1}^{\;\prime\,2}
\vec{q}_{2}^{\;2}}
{\vec{k}^{\,2}\vec{q}^{\;2}}\right)+\frac{1}{\vec{q}_2^{\,\,2}}
\ln\left(\frac{\vec{q}_{2}^{\;\prime\,2}\vec{q}_{1}^{\;2}}
{\vec{k}^{\,2}\vec{q}^{\;2}}\right)+\frac{1}{\vec{k}^{\,2}}\ln\left(\frac
{\vec{q}_{1}^{\;\prime\,2}\vec{q}_{2}^{\;\prime\,2}}{\vec{q}_{1}^{\;2}
\vec{q}_{2}^{\;2}}\right)
\]
\begin{equation}
 -2\frac{\vec q_1\vec k}{\vec{k}^{\,2}\vec{q}_1^{\,\,2}}
\ln\left(\frac{\vec{q}_{1}^{\;\prime\,2}}
{\vec{k}^{\,2}}\right)
+2\frac{\vec q_2\vec k}{\vec{k}^{\,2}\vec{q}_2^{\,\,2}}
\ln\left(\frac{\vec{q}_{2}^{\;\prime\,2}}
{\vec{k}^{\,2}}\right)
-2\frac{\vec q_1 \vec{q}_2}{\vec{q}_1^{\,\,2}\vec{q}_2^{\,\,2}}\ln
\left(\frac{\vec{q}^{\;2}}{\vec{k}^{\,2}}\right)~. \label{R u momentum}
\end{equation}
Note that $R_u$ has the same gauge invariance properties as $R$:
\[
R_u(\vec{q}_{1},\vec{q}_{2};  \vec{q}_1)=R_u(\vec{q}_{1},\vec{q}_{2};
-\vec{q}_2)=0, \;\;
\]
\begin{equation}
(\vec{q}_{1}^{\;2}\vec{q}_{2}^{\;2}R_u(\vec{q}_{1},\vec{q}_{2};
\vec{k}))|_{\vec{q}_{1}=0}=(\vec{q}_{1}^{\;2}\vec{q}_{2}^{\;2}
R_u(\vec{q}_{1},\vec{q}_{2};  \vec{k}))|_{\vec{q}_{2}=0}=0~.
\label{g i for R u}
\end{equation}
Indeed, these properties are required to conserve the gauge invariance in the
transformation~(\ref{kernel transformation}).

Another important property of $R_u$ is the absence of either infrared,
or  ultraviolet non-integrable singularities, thus leading to
convergence of the integral
\begin{equation}
\int\frac{d\vec{k}_1d\vec{k}_2}{\pi} \delta(\vec{k}-\vec{k}_1-\vec{k}_2)
R_u(\vec{q}_1-\vec{k}_1, \vec{q}_2+\vec{k}_1; \vec{k}_2)=
-\ln\left(\frac{\vec{q}_{1}^{\;\prime\,2}}{\vec{q}^{\;2}}\right)
\ln\left(  \frac{\vec{q}_{2}^{\;\prime\,2}}{\vec{q}^{\;2}}\right)~.
\label{int R u}
\end{equation}
The calculation of this integral and of the integrals appearing below is
described in the Appendix. The result~(\ref{int R u}) follows
from~(\ref{int phi ln}) and~(\ref{int phi a-l b-l}) with a subsequent
elementary integration over $l$.

Having Eqs.~(\ref{K B  momentum}) and~(\ref{U momentum}), it is quite
straightforward to write the commutator $\left[\hat{\mathcal{K}}^{(B)},\hat
{U}\right]$ in the form
\[
\langle\vec{q}_{1},\vec{q}_{2}|\alpha_s\left[\hat{\mathcal{K}}^{(B)},\hat
{U}\right]|\vec{q}_{1}^{\,\,\prime}%
,\vec{q}_{2}^{\,\,\prime}\rangle=\delta(\vec{q}_{11^{\prime}}+\vec
{q}_{22^{\prime}})\frac{\alpha^2_s N^2_c}{8\pi^{3}}\left[ \frac{\beta_0}{2N_c}
\ln\left(\frac{\vec{q}_1^{\;2}\vec{q}_2^{\;2}}{\vec{q}_{1}^{\;\prime\,2}
\vec{q}_{2}^{\;2}}
\right) R(\vec{q}_1, \vec{q}_2; \vec{k})\right.
\]
\begin{equation}
\left.+ \int\frac{d\vec{l}}{\pi}\left(V(\vec{q}^{\;\prime}_1,
\vec{q}^{\;\prime}_2; \vec{l})-V(\vec{q}_1, \vec{q}_2; \vec{l})\right)\;
R_u(\vec{q}_1, \vec{q}_2; \vec{k})+
F(\vec{q}_1, \vec{q}_2; \vec{k})\right]~,\label{[] as sum}
\end{equation}
where
\[
F(\vec{q}_1, \vec{q}_2; \vec{k})=\int\frac{d\vec{k}_1d\vec{k}_2}{\pi}
\delta(\vec{k}-\vec{k}_1-\vec{k}_2){\cal F}(\vec{q}_1, \vec{q}_2; \vec{k}_1,
\vec{k}_2)~,\;\;{\cal F}(\vec{q}_1, \vec{q}_2; \vec{k}_1, \vec{k}_2)=
\]
\begin{equation}
 =R(\vec{q}_1, \vec{q}_2; \vec{k}_1)R_u(\vec{q}_1-\vec{k}_1, \vec{q}_2
+\vec{k}_1; \vec{k}_{2})-R_u(\vec{q}_1, \vec{q}_2; \vec{k}_1)R(\vec{q}_1
-\vec{k}_1, \vec{q}_2+\vec{k}_1; \vec{k}_{2})~. \label{F definition}
\end{equation}
The infrared divergent pieces in the virtual parts entering the integral
over $d\vec l$ in Eq.~(\ref{[] as sum}) cancel, and one can easily obtain
(see~(\ref{int v}))
\begin{equation}
\int\frac{d\vec{l}}{\pi}\left(V(\vec{q}^{\;\prime}_1, \vec{q}^{\;\prime}_2;
\vec{l})-V(\vec{q}_1, \vec{q}_2; \vec{l})\right) =
\ln\left(\frac{\vec{q}_{1}^{\;\prime\,2}\vec{q}_{2}^{\;\prime\,2}}
{\vec{q}_1^{\;2}\vec{q}_2^{\;2}}\right)~. \label{virtual contribution}
\end{equation}
Unfortunately, the calculation of $F(\vec{q}_1, \vec{q}_2; \vec{k})$ is not
so easy, both because of the presence of a great number of terms in
${\cal F}(\vec{q}_1, \vec{q}_2; \vec{k}_1, \vec{k}_2)$ and of the complexity
of the integration. One of the reasons of this complexity is the singularity
of $R(\vec{q}_1, \vec{q}_2; \vec{k})$ at $\vec{k}^{\,2}=0$. Of course, this
singularity disappears in $F(\vec{q}_1, \vec{q}_2;\vec{k})$, 
Eq.~(\ref{F definition}). To make this evident, let us write
\begin{equation}
R(\vec{q}_1, \vec{q}_2; \vec{k})=\frac{2}{\vec{k}^{\,2}}
+R_f(\vec{q}_1, \vec{q}_2; \vec{k}), \;\; R_f(\vec{q}_{1},\vec{q}_{2};
\vec{k})= -2\frac{\vec{q}_{1}\vec{k}}{\vec{k}^{\,2}\vec{q}%
_{1}^{\,\,2}}+2\frac{\vec{q}_{2}\vec{k}}{\vec{k}^{\,2}\vec{q}_{2}^{\,\,2}%
} -2\frac{\vec{q}_{1}\vec{q}_{2}}{\vec{q}_{1}^{\,\,2}\vec{q}_{2}%
^{\,\,2}}\;,\label{R s  momentum}
\end{equation}
and divide $F(\vec{q}_1, \vec{q}_2; \vec{k})$, Eq.~(\ref{F definition}), 
into three pieces:
\begin{equation}
F(\vec{q}_1, \vec{q}_2; \vec{k}) =\sum_{i=1}^3F_i(\vec{q}_1, \vec{q}_2;
\vec{k}),
\label{F division}
\end{equation}
where
\begin{equation}
F_1(\vec{q}_1, \vec{q}_2; \vec{k})=\int\frac{d\vec{k}_1}{\pi}R_f(\vec{q}_1,
\vec{q}_2; \vec{k}_1)R_u(\vec{q}_1-\vec{k}_1, \vec{q}_2+\vec{k}_1;
\vec{k}-\vec{k}_{1})\;,
\label{F1}
\end{equation}
\begin{equation}
F_2(\vec{q}_1, \vec{q}_2; \vec{k})=-\int\frac{d\vec{k}_1}{\pi}R_u(\vec{q}_1,
\vec{q}_2; \vec{k}_1)R_f(\vec{q}_1-\vec{k}_1, \vec{q}_2+\vec{k}_1;
\vec{k}-\vec{k}_{1})\;,
\label{F2}
\end{equation}
\begin{equation}
F_3(\vec{q}_1, \vec{q}_2; \vec{k})=\int\frac{d\vec{k}_1}{\pi}\frac{2}
{\vec{k}_1^{\,2}}\left(R_u(\vec{q}_1-\vec{k}_1, \vec{q}_2+\vec{k}_1;
\vec{k}-\vec{k}_{1})-R_u(\vec{q}_1, \vec{q}_2; \vec{k}-\vec{k}_1)\right)\;.
\label{F3}
\end{equation}
Now all the three pieces have no infrared singularities, the first two of them
because of the absence of singularities in the integrands, and the last one 
because of the evident cancellation between the two terms with $R_u$ in
Eq.~(\ref{F3}) at $\vec{k}_1=0$. The integration of the first piece can
be performed with the help of 
Eqs.~(\ref{int R u}),~(\ref{int vl2}),~(\ref{int vl3}) and~(\ref{int vl30}) 
and gives
\[
F_1(\vec{q}_{1}, \vec{q}_{2}; \vec{k})=
\left(\frac{\vec{q}_{1}\vec{q}_{2}}{\vec{q}_{1}^{\;2}\vec{q}_{2}^{\;2}}
-\frac{1}{\vec{q}_{1}^{\;2}}\right)\ln\left(\frac{\vec{q}^{\;2}\vec{q}_1^{\;2}}
{\vec{q}_{2}^{\;\prime\,2}\vec{k}^{\,2}}\right)
\ln\left(\frac{\vec{q}_{1}^{\;\prime\,2}\vec{q}_{2}^{\;2}}{\vec{q}^{\;2}
\vec{k}^{\,2}}\right)
+\frac{\vec{q}_{1}\vec{q}_{2}}{\vec{q}_{1}^{\;2}\vec{q}_{2}^{\;2}}
\ln\left(\frac{\vec{q}_1^{\;\prime\,2}}
{\vec{q}^{\;2}}\right)
\ln\left(\frac{\vec{q}_{2}^{\;\prime\,2}}
{\vec{q}^{\;2}}\right)
\]
\[
+4\left(\frac{[\vec{q}_{2}\times\vec{k}]}{\vec{q}_{2}^{\;2}\vec{k}^{\,2}}-
\frac{[\vec{q}_{1}\times\vec{k}]}{\vec{q}_{1}^{\;2}\vec{k}^{\,2}}\right)
[\vec{q}_{1}\times\vec{k}]
I_{\vec{k}, \vec{q}_{1}^{\;\prime}} +2\frac{[\vec{q}_{1}\times\vec{q}_{2}]^2}
{\vec{q}_{1}^{\;2}\vec{q}_{2}^{\;2}}I_{\vec{q}_{1}, \vec{q}_{2}}
\]
\begin{equation}
+2\frac{[\vec{q}_{1}\times\vec{q}_{2}][\vec{q}_{1}^{\;\prime}
\times\vec{q}_{2}^{\;\prime}]}
{\vec{q}_{1}^{\;2}\vec{q}_{2}^{\;2}}I_{\vec{q}_{1}^{\;\prime},
\vec{q}_{2}^{\;\prime}}+  \vec{q}_{1}\leftrightarrow -\vec{q}_{2}\;.
\label{F1 final}
\end{equation}
Here
\begin{equation}
I_{\vec p, \vec q}=
\int_{0}^{1}\frac{dx}{(\vec p +x\vec q)^{2}}\ln\left(\frac{\vec p^{\;2}}
{x^2\vec q^{\;2}}\right)~ \label{I p q 1}
\end{equation}
is the di-logarithmic function with high symmetry,
\begin{equation}
I_{\vec p,\vec q}=I_{-\vec p,-\vec q}=I_{\vec q, \vec p}=I_{\vec p,-\vec p
-\vec q}~.
\label{I definition}
\end{equation}
The representation exhibiting these properties~\cite{Fadin:2002hz} is
\begin{equation}
I_{\vec p,\vec q}=\int_{0}^{1}\int_{0}^{1}\int_{0}^{1}\frac{dx_{1}dx_{2}dx_{3}
\delta(1-x_{1}-x_{2}-x_{3})}{(\vec p^{\;2}x_{1}+\vec q^{\;2}x_{2}+(\vec p
+\vec q)^{2}x_{3})(x_{1}x_{2}+x_{1}x_{3}+x_{2}x_{3})}~.\label{I symmetric}
\end{equation}
Other useful representations are
\[
I_{\vec p,\vec q}=\int_{0}^{1}
\frac{dx}{a(1-x)+bx-c x(1-x)}\ln\left(  \frac{a(1-x)+bx}{cx(1-x)}\right)
\]
\begin{equation}
=\int_{0}^{1}dx\int_{0}^{1}{dz}\;\frac{1}{cx(1-x)z+(b(1-x)+ax)(1-z)}~,
\label{integral I}
\end{equation}
where $a=\vec p^{\;2}, b=\vec q^{\;2}, c = (\vec p+\vec q)^{2} $.

Note that $F_1$ must turn into zero at $\vec{q}_{1}^{\;\prime}=0$ or
$\vec{q}_{2}^{\;\prime}=0$  due to the gauge invariance of $R_u$. It is easy
to see from Eq.~(\ref{F1 final}) that this property is fulfilled.

Unfortunately, neither $F_2$ nor $F_3$ possess such property. Moreover, the
separation~(\ref{F division}) destroys the good behaviour of $R(\vec{q}_1,
\vec{q}_2; \vec{k})$ in the ultraviolet region, so that the
integrals~(\ref{F2})  and~(\ref{F3})  diverge at large $\vec{k}_1^{\,2}$
and we have to introduce an ultraviolet cut-off $\Lambda^2$ for them.
The loss of gauge invariance and ultraviolet convergence of the integrals
makes them more complex. Using~(\ref{int vl3})--(\ref{int v3l0}) we obtain
\[
F_2(\vec{q}_{1}, \vec{q}_{2}; \vec{k})=
\frac{1}{\vec{q}_{1}^{\;2}}\left(\ln\left(\frac{\vec{q}_2^{\;2}}
{\vec{q}^{\;2}}\right)
\ln\left(\frac{\Lambda^4\vec{q}^{\;4}\vec{q}_{1}^{\;2}}
{\vec{q}_{1}^{\;\prime\,6}
\vec{q}_{2}^{\;\prime\,2}\vec{q}_2^{\;2}}\right)
+\ln\left(\frac{\vec{q}_1^{\;\prime\,2}}
{\vec{k}^{\,2}}\right)\ln\left(\frac{\vec{q}_{1}^{\;2}\vec{q}_{2}^{\;2}}
{\vec{k}^{\,2}
\vec{q}_{2}^{\;\prime\,2}}\right) \right)+\frac{\vec{q}_{1}\vec{q}_{2}}
{\vec{q}_{1}^{\;2}\vec{q}_{2}^{\;2}}
\]
\[
\times\left(\ln^2\left(\frac{\Lambda^2}
{\vec{q}^{\;2}}\right)-\ln\left(\frac{\vec{q}_1^{\;2}}
{\vec{q}^{\;2}}\right)\ln\left(\frac{\vec{q}_{2}^{\;2}}{\vec{q}^{\;2}}\right)
-\ln\left(\frac{\vec{q}_1^{\;\prime\,2}}
{\vec{q}^{\;2}}\right)\ln\left(\frac{\vec{q}_{1}^{\;\prime\,2}
\vec{q}_{2}^{\;\prime\,2}}
{\vec{q}^{\;4}}\right)
-\ln\left(\frac{\vec{k}^{\,2}}{\vec{q}_1^{\;\prime\,2}}\right)
\ln\left(\frac{\vec{k}^{\,2}\vec{q}_1^{\;\prime\,2}}{\vec{q}_1^{\;4}}
\right)\right)
\]
\[
+2\frac{\vec{q}_{1}\vec{k}}{\vec{q}_{1}^{\;2}\vec{k}^{\,2}}
\ln\left(\frac{\vec{q}^{\;2}}
{\vec{q}_2^{\;2}}\right)\ln\left(\frac{\vec{q}_{1}^{\;2}
\vec{q}_{2}^{\;\prime\,2}}
{\vec{q}_{1}^{\;\prime\,2}\vec{q}_{2}^{\;2}}\right)+4\left(\frac{[\vec{q}_{1}
\times\vec{k}]}
{\vec{q}_{1}^{\;2}\vec{k}^{\,2}}-
\frac{[\vec{q}_{2}\times\vec{k}]}{\vec{q}_{2}^{\;2}\vec{k}^{\,2}}-
\frac{[\vec{q}_{1}\times\vec{q}_2]}{\vec{q}_{2}^{\;2}\vec{q}_2^{\;2}}\right)
[\vec{q}_{1}\times\vec{k}]
I_{\vec{k}, \vec{q}_{1}^{\;\prime}}
\]
\begin{equation}
-2\left(\frac{[\vec{q}_{1}\times\vec{k}]}{\vec{q}_{1}^{\;2}\vec{k}^{\,2}}+
\frac{[\vec{q}_{2}\times\vec{k}]}{\vec{q}_{2}^{\;2}\vec{k}^{\,2}}+
\frac{[\vec{q}_{1}\times\vec{q}_2]}{\vec{q}_{2}^{\;2}\vec{q}_2^{\;2}}\right)
\Biggl([\vec{q}_{1}\times\vec{q}_{2}]I_{\vec{q}_{1}, \vec{q}_{2}}
-[\vec{q}_{1}^{\;\prime}\times\vec{q}_{2}^{\;\prime}]
I_{\vec{q}_{1}^{\;\prime}, \vec{q}_{2}^{\;\prime}}\Biggr)+  \vec{q}_{1}
\leftrightarrow -\vec{q}_{2}~.  \label{F2 final}
\end{equation}
The result for $F_3(\vec{q}_1,\vec{q}_2; \vec{k})$ can be obtained
using Eqs.~(\ref{int vl2}),~(\ref{int s3l0})--(\ref{int s3ld}) and reads
\[
F_3(\vec{q}_{1}, \vec{q}_{2}; \vec{k})=
\frac{1}{\vec{q}_{1}^{\;2}}\left(\ln\left(\frac{\vec{q}^{\;2}}
{\vec{q}_2^{\;2}}\right)
\ln\left(\frac{\Lambda^4\vec{q}^{\;2}}{\vec{q}_{1}^{\;4}
\vec{q}_{2}^{\;2}}\right)-2\ln\left(\frac{\vec{q}_1^{\;\prime\,2}}
{\vec{k}^{\,2}}\right)\ln\left(\frac{\vec{q}_{1}^{\;\prime\,2}}
{\vec{q}_1^{\;2}}\right) \right)-\frac{\vec{q}_{1}\vec{q}_{2}}
{\vec{q}_{1}^{\;2}\vec{q}_{2}^{\;2}}
\]
\[
\times\left(\ln^2\left(\frac{\Lambda^2}
{\vec{q}^{\;2}}\right)-2\ln^2\left(\frac{\vec{q}_1^{\;2}}
{\vec{q}^{\;2}}\right)
-\ln\left(\frac{\vec{q}_1^{\;\prime\,2}}
{\vec{q}^{\;2}}\right)\ln\left(\frac{\vec{q}_{2}^{\;\prime\,2}}
{\vec{q}^{\;2}}\right)
+2\ln\left(\frac{\vec{k}^{\,2}}{\vec{q}_1^{\;2}}\right)
\ln\left(\frac{\vec{q}_1^{\;\prime\,2}}{\vec{q}_1^{\;2}}
\right)\right)
\]
\[
-4\frac{\vec{q}_{1}\vec{k}}{\vec{q}_{1}^{\;2}\vec{k}^{\,2}}
\ln\left(\frac{\vec{k}^{\,2}}
{\vec{q}_1^{\;\prime\,2}}\right)
\ln\left(\frac{\vec{q}_1^{\;\prime\,2}}{\vec{q}_1^{\;2}}
\right)-\frac{2}{\vec k^{\,2}}\ln^2\left(\frac{\vec{q}_1^{\;\prime\,2}}
{\vec{q}_1^{\;2}}
\right)
\]
\begin{equation}
+2\frac{[\vec{q}_{1}\times\vec{q}_2]}{\vec{q}_{1}^{\;2}\vec{q}_2^{\;2}}
\Biggl(2[\vec{q}_{1}\times\vec{k}]I_{\vec{k}_{1}, \vec{q}^{\;\prime}_{1}}
-[\vec{q}_{1}^{\;\prime}\times\vec{q}_{2}^{\;\prime}]
I_{\vec{q}_{1}^{\;\prime}, \vec{q}_{2}^{\;\prime}}\Biggr)
+ \vec{q}_{1}\leftrightarrow -\vec{q}_{2}\;.  \label{F3 final}
\end{equation}
From the Eq.~(\ref{F division}) and the definitions~(\ref{F1})--(\ref{F3}) 
it follows
\[
F(\vec{q}_{1}, \vec{q}_{2}; \vec{k})=
\frac{2}{\vec{q}_{1}^{\;2}}\ln\left(\frac{\vec{q}_1^{\;2}}
{\vec{q}_1^{\;\prime\,2}}\right)
\ln\left(\frac{\vec{q}_{1}^{\;\prime\,2}\vec{q}_{2}^{\;2}}{
\vec{q}^{\;2}\vec{k}^{\,2}}\right)+\frac{2\vec{q}_{1}\vec{q}_{2}}
{\vec{q}_{1}^{\;2}\vec{q}_{2}^{\;2}}\ln\left(\frac{\vec{q}_1^{\;2}}
{\vec{q}_1^{\;\prime\,2}}\right)
\ln\left(\frac{\vec{k}^{\,2}}{
\vec{q}^{\;2}}\right)
\]
\[
+2\frac{\vec{q}_{1}\vec{k}}{\vec{q}_{1}^{\;2}\vec{k}^{\,2}}
\left(\ln\left(\frac{\vec{q}^{\;2}}
{\vec{q}_2^{\;2}}\right)\ln\left(\frac{\vec{q}_{1}^{\;2}
\vec{q}_{2}^{\;\prime\,2}}
{\vec{q}_{1}^{\;\prime\,2}\vec{q}_{2}^{\;2}}\right)
+2\ln\left(\frac{\vec{q}_{1}^{\;\prime\,2}}
{\vec{q}_1^{\;2}}\right)\ln\left(\frac{\vec{q}_{1}^{\;\prime\,2}}
{\vec{k}^{\,2}}\right)\right) -\frac{2}{\vec k^{\,2}}
\ln^2\left(\frac{\vec{q}_1^{\;\prime\,2}}{\vec{q}_1^{\;2}}
\right)
\]
\begin{equation}
-2\left(\frac{[\vec{q}_{1}\times\vec{q}_{2}]}{\vec{q}_{1}^{\;2}
\vec{q}_{2}^{\;2}}
+2\frac{[\vec{q}_{1}\times\vec{k}]}{\vec{q}_{1}^{\;2}\vec{k}^{\,2}}
\right)\left([\vec{q}_{1}\times\vec{q}_{2}]I_{\vec{q}_{1}, \vec{q}_{2}}
-[\vec{q}_{1}^{\;\prime}\times\vec{q}_{2}^{\;\prime}]I_{\vec{q}_{1}^{\;\prime},
\vec{q}_{2}^{\;\prime}}\right) +\vec q_1\leftrightarrow -\vec q_2~.
\label{F final}
\end{equation}
The definition~(\ref{F definition}) and the properties~(\ref{g i for R})
and~(\ref{g i for R u}) of $R$ and $R_u$, respectively, secure the gauge
invariance of $F$:
\[
F(\vec{q}_{1}, \vec{q}_{2}; \vec{q}_1)=F(\vec{q}_{1}, \vec{q}_{2}; -\vec{q}_2)
=0,
\]
\begin{equation}
( \vec{q}_{1}^{\;2}\vec{q}_{2}^{\;2} F(\vec{q}_{1},\vec{q}_{2};
\vec{q}_1))|_{\vec{q}_{1}=0}=(\vec{q}_{1}^{\;2}\vec{q}_{2}^{\;2}
F(\vec{q}_{1},\vec{q}_{2};  \vec{q}_1))|_{\vec{q}_{2}=0}=0~.
\label{g i for F}
\end{equation}
The fulfilment of these properties can be easily seen from Eq.~(\ref{F final}).

Finally, Eq.~(\ref{[] as sum}) together with
Eqs.~(\ref{R u momentum}),~(\ref{virtual contribution})
and~(\ref{F final}) gives
\[
\langle\vec{q}_{1},\vec{q}_{2}|\alpha_{s}[{\hat{\mathcal{K}}}^{(B)},
{\hat{U}}]|\vec{q}_{1}^{\,\,\prime
},\vec{q}_{2}^{\,\,\prime}\rangle=\delta(\vec{q}_{11^{\prime}}+\vec
{q}_{22^{\prime}})\frac{\alpha^2_{s}N^2_{c}}{8\pi^{3}}\left[ -\frac{\beta_0}
{2N_c}
R(\vec{q}_{1}, \vec{q}_{2}; \vec{k})\ln\left(  \frac{\vec{q}_{1}^{\;\prime\,2}
\vec{q}_{2}^{\;\prime\,2}}%
{\vec{q}_{1}^{\;2}\vec{q}_{2}^{\;2}}\right) +
\right.
\]
\[
\left. + \frac{\vec{q}_{1}^{\;\prime\,2}}{\vec{q}_{1}^{\;2}\vec{k}^{\,2}}
\ln\left(\frac{\vec{q}_{1}^{\;2}\vec{q}_{2}^{\;\prime\,2}}{\vec{q}_{2}^{\;2}
\vec{q}_{1}^{\;\prime\,2}}\right)
\ln\left(\frac{\vec{q}_{2}^{\;2}\vec{q}_{1}^{\;\prime\,2}}{\vec{q}^{\;2}
\vec{k}^{\,2}}
\right) +\frac{\vec{q}_{2}^{\;\prime\,2}}{\vec{q}_{2}^{\;2}\vec{k}^{\,2}}
\ln\left(\frac{\vec{q}_{2}^{\;2}\vec{q}_{1}^{\;\prime\,2}}{\vec{q}_{1}^{\;2}
\vec{q}_{2}^{\;\prime\,2}}\right)
\ln\left(\frac{\vec{q}_{1}^{\;2}\vec{q}_{2}^{\;\prime\,2}}{\vec{q}^{\;2}
\vec{k}^{\,2}}
\right)\right.
\]
\begin{equation}
\left. -4\left(\frac{[\vec{q}_{1}\times\vec{q}_{2}]}{\vec{q}_{1}^{\;2}
\vec{q}_{2}^{\;2}}
+\frac{[\vec{q}_{1}\times\vec{k}]}{\vec{q}_{1}^{\;2}\vec{k}^{\,2}}
+\frac{[\vec{q}_{2}\times\vec{k}]}{\vec{q}_{2}^{\;2}\vec{k}^{\,2}}
\right)\left([\vec{q}_{1}\times\vec{q}_{2}]I_{\vec{q}_{1}, \vec{q}_{2}}
-[\vec{q}_{1}^{\;\prime}\times\vec{q}_{2}^{\;\prime}]I_{\vec{q}_{1}^{\;\prime},
\vec{q}_{2}^{\;\prime}}\right) \right]. \label{final commutator}
\end{equation}

\section{Use of M\"{o}bius space}
\label{sec:Mobius}

Since the result~(\ref{final commutator}) was derived by means of
lengthy and intricate calculations, we  want to obtain  it in a quite
independent way, starting from the M\"{o}bius forms of the kernel
${\hat{\cal K}}^{(B)}$ and of the operator $\hat{U}$, calculating their
commutator and restoring the complete commutator~(\ref{final commutator}) in
the momentum space from its M\"{o}bius form. Simultaneously, the efficiency
of the method of restoration developed in Ref.~\cite{Fadin:2011jg}
will be  demonstrated. Here this alternative derivation is illustrated.

As it is known~\cite{Fadin:2006ha}, the M\"{o}bius form of the kernel
${\hat{\cal K}}^{(B)}$ coincides with the kernel of the colour dipole
model~\cite{dipole} and can be written as
\[
\langle\vec{r}_{1}\vec{r}_{2}|{\hat{\cal K}}^{(B)}_{M}|\vec{r}_{1}^{\;\prime}
\vec{r}%
_{2}^{\;\prime}\rangle=\frac{\alpha_{s}N_{c}}{2\pi^{2}}\int d\vec{r
}_0g(\vec{r}_{1}, \vec{r}_{2},\vec{r}_{0})
\]
\begin{equation}
\times \Biggl[\delta(\vec{r}_{11^{\prime}})\delta(\vec
{r}_{2^{\prime}0})+\delta(\vec{r}_{1^{\prime}0})\delta(\vec
{r}_{22^{\prime}})-\delta(\vec{r}_{11^{\prime}})\delta({r}_{22^{\prime}%
})\Biggr]~, \label{K B M}%
\end{equation}
where
\begin{equation}
g(\vec{r}_{1}, \vec{r}_{2},\vec{r}_{0})=g(\vec{r}_{2}, \vec{r}_{1},
\vec{r}_{0})=
\frac{\vec{r}_{12}^{\;2}}{\vec{r}_{10}^{\,\,2}\vec{r}_{20}^{\,\,2}}\;.
\label{g function}
\end{equation}
The M\"{o}bius form of the operator $U$ was found in Ref.~\cite{Fadin:2011jg}.
Omitting the term with $\hat{\cal K}^{(B)}$, which does not contribute to the
commutator in~(\ref{QC kernel}), one has
\[
\langle\vec{r}_{1}\vec{r}_{2}|\alpha_s\hat{U}_M|\vec{r}_{1}^{\;\prime}%
\vec{r}_{2}^{\;\prime}\rangle=\frac{\alpha_{s}N_{c}}{4\pi^{2}}
\Biggl[\delta(\vec{r}
_{11^{\prime}})V_1(\vec{r}_{1}, \vec{r}_{2},\vec{r}_{2}^{\;\prime})
+\delta(\vec{r}_{22^{\prime}})V_1(\vec{r}_{2}, \vec{r}_{1},
\vec{r}_{1}^{\;\prime})
\]
\begin{equation}
+\frac{1}{\pi}V_3(\vec{r}_{1}, \vec{r}_{2},\vec{r}_{1}^{\;\prime},
\vec{r}_{2}^{\;\prime})\Biggr]+\frac{\alpha_{s}\beta_{0}}{8\pi^{2}}
\Biggl[\delta(\vec{r}_{11^{\prime}})\left(\frac{1}{\vec{r}_{22^{\prime}}
^{\,\,2}}-\frac{1}{\vec{r}_{12^{\prime}}
^{\,\,2}} \right) +\delta(\vec{r}_{22^{\prime}})
\left(\frac{1}{\vec{r}_{11^{\prime}}
^{\,\,2}}-\frac{1}{\vec{r}_{21^{\prime}}
^{\,\,2}} \right)\Biggr]~,
\label{U coordinate}
\end{equation}
where
\begin{equation}
V_1(\vec{r}_{1}, \vec{r}_{2},\vec{r}_{2}^{\;\prime})
=\frac{\vec{r}_{12}^{\,\,2}}{\vec{r}_{12'}^{\,\,2}\vec{r}
_{22'}^{\,\,2}}\ln\left(\frac{\vec{r}_{12}^{\,\,2}}{\vec{r}_{22'}^{\,\,2}}
\right) +\frac{1}{\vec{r}_{22'}^{\,\,2}}\ln\left(  \frac{\vec
{r}_{22'}^{\,\,2}}{\vec{r}_{12'}^{\,\,2}}\right), \label{V1}
\end{equation}
\begin{equation}
V_3(\vec{r}_{1}, \vec{r}_{2},\vec{r}_{1}^{\;\prime}, \vec{r}_{2}^{\;\prime})
=V_3(\vec{r}_{2}, \vec{r}_{1},\vec{r}_{2}^{\;\prime}, \vec{r}_{1}^{\;\prime})
= \frac{1}{\vec {r}_{1^{\prime}2^{\prime}}^{\,\, 2}}
\left[ \frac{2\vec{r}_{1 1^{\prime}}\vec{r}_{2 2^{\prime}}}
{\vec{r}_{1 1^{\prime}}^{\,\, 2}\vec{r}_{2 2^{\prime}}^{\,\, 2}}
- \frac{\vec{r}_{1 1^{\prime}}\vec{r}_{1 2^{\prime}}}
{\vec{r}_{1 1^{\prime}}^{\,\, 2}\vec{r}_{1 2^{\prime}}^{\,\, 2}}
- \frac{\vec{r}_{2 1^{\prime}}\vec{r}_{2 2^{\prime}}}
{\vec{r}_{2 1^{\prime}}^{\,\, 2}\vec{r}_{2 2^{\prime}}^{\,\, 2}}\right]~.
\label{V3}
\end{equation}
The treatment  of the term with $\beta_0$ in $\hat{U}$ can be performed
quite easily in the momentum space (see Eq.~(\ref{[] as sum})), so that in the
following we will omit this term, denoting the remaining part of $\hat{U}$
as $\hat{U}^s$. With the notation~(\ref{K B M})--(\ref{V3}) the M\"{o}bius
form for the commutator $[\hat{\mathcal{K}}^{(B)},\hat{U}^s]$ can be presented
as
\[
\langle\vec{r}_{1}\vec{r}_{2}|\alpha_s[\hat{\mathcal{K}}^{(B)},\hat
{U}^s]_M|\vec{r}_{1}^{\;\prime}\vec{r}_{2}^{\;\prime}\rangle
=\frac{\alpha_s^2N_c^2}{8\pi^3}\left[\delta(\vec{r}_{11^{\prime}})
J(\vec{r}_{1}, \vec{r}_{2},\vec{r}_{2}^{\;\prime})+\frac{1}{\pi}
F(\vec{r}_{1}, \vec{r}_{2},\vec{r}_{1}^{\;\prime}, \vec{r}_{2}^{\;\prime})
\right.
\]
\begin{equation}
\left.
\frac{1}{\pi}I(\vec{r}_{1}, \vec{r}_{2},\vec{r}_{1}^{\;\prime},
\vec{r}_{2}^{\;\prime})
\right] +1\leftrightarrow 2~, \label{mebius commutator}
\end{equation}
where $1\leftrightarrow 2$ means the substitution $\vec{r}_{1}\leftrightarrow
\vec{r}_{2}, \; \vec{r}_{1}^{\;\prime}\leftrightarrow \vec{r}_{2}^{\;\prime}$.
The first two terms in the square brackets in Eq.~(\ref{mebius commutator}) 
come from the term with $V_1(\vec{r}_{1}, \vec{r}_{2},\vec{r}_{2}^{\;\prime})$
in  Eq.~(\ref{U coordinate})  and are written as
\[
J(\vec{r}_{1}, \vec{r}_{2},\vec{r}_{2}^{\;\prime}) =\int \frac{d\x_0}{\pi}
\left[g(\vec{r}_{1}, \vec{r}_{2},\vec{r}_{0})V_1(\vec{r}_{1}, \vec{r}_{0},
\vec{r}_{2}^{\;\prime}) -V_1(\vec{r}_{1}, \vec{r}_{2},\vec{r}_{0})
g(\vec{r}_{1}, \vec{r}_{0},\vec{r}_{2}^{\;\prime})\right.
\]
\begin{equation}
\left.-(g(\vec{r}_{1}, \vec{r}_{2},\vec{r}_{0})-g(\vec{r}_{1},
\vec{r}_{2}^{\;\prime}, \vec{r}_{0}))V_1(\vec{r}_{1},  \vec{r}_{2},
\vec{r}_{2}^{\;\prime})\right] \label{J definition}
\end{equation}
and
\begin{equation}
F(\vec{r}_{1}, \vec{r}_{2},\vec{r}_{1}^{\;\prime}, \vec{r}_{2}^{\;\prime})=
g(\vec{r}_{2}, \vec{r}_{1},\vec{r}_{1}^{\;\prime})V_1(\vec{r}_{1}^{\;\prime},
\vec{r}_{2},\vec{r}_{2}^{\;\prime})-V_1(\vec{r}_{1}, \vec{r}_{2},
\vec{r}_{2}^{\;\prime})
g(\vec{r}_{2}^{\;\prime}, \vec{r}_{1},\vec{r}_{1}^{\;\prime})\;.
\label{F coordinate}
\end{equation}
The last term in the square brackets in Eq.~(\ref{mebius commutator}) related
with $V_3$ is presented in the form
\[
I(\vec{r}_{1}, \vec{r}_{2},\vec{r}_{1}^{\;\prime}, \vec{r}_{2}^{\;\prime})
=\int \frac{d\x_0}{\pi}
\left[\left(\frac{1}{\xs_{10}}-\frac{\x_{10}\x_{20}}{\xs_{10}\xs_{20}}\right)
V_3(\vec{r}_{1}, \vec{r}_{0},\vec{r}_{1}^{\;\prime}, \vec{r}_{2}^{\;\prime})
+\left(\frac{1}{\xs_{20}}-\frac{\x_{10}\x_{20}}{\xs_{10}\xs_{20}}\right)
\right.
\]
\[
\times \Biggl(
V_3(\vec{r}_{1}, \vec{r}_{0},\vec{r}_{1}^{\;\prime}, \vec{r}_{2}^{\;\prime})-
V_3(\vec{r}_{1}, \vec{r}_{2},\vec{r}_{1}^{\;\prime}, \vec{r}_{2}^{\;\prime})
\Biggr)
-\frac{1}{\xs_{02'}}\Biggl(
V_3(\vec{r}_{1}, \vec{r}_{2},\vec{r}_{1}^{\;\prime}, \vec{r}_{0})
\frac{\xs_{01'}}{\xs_{1'2'}}
\]
\begin{equation}
\left.
-V_3(\vec{r}_{1}, \vec{r}_{2},\vec{r}_{1}^{\;\prime}, \vec{r}_{2}^{\;\prime})
\Biggr)
-\frac{\x_{1'0}\x_{2'0}}{\xs_{1'0}\xs_{2'0}}V_3(\vec{r}_{1}, \vec{r}_{2},
\vec{r}_{1}^{\;\prime}, \vec{r}_{2}^{\;\prime})\right] ~.
\label{I definition bis}
\end{equation}
Note that each of the $J, F, I$ functions independently turns into zero 
at $\vec{r}_{12}=0$.
In contrast to the function $F$, which is given explicitly
by Eq.~(\ref{F coordinate}), the functions $J$ and $I$ are expressed in terms 
of the integrals~(\ref{J definition}) and~(\ref{I definition bis}), 
respectively.
The integrals are not very intricate, although their calculation is
complicated by the ultraviolet divergences existing in separate terms.
The integrands in~(\ref{J definition}) and~(\ref{I definition bis}) are written in
such a way so as to make the cancellation evident. The results of the
integration (which can be performed by the method described in the Appendix)
are very simple:
\begin{equation}
J(\vec{r}_{1}, \vec{r}_{2},\vec{r}_{2}^{\;\prime}) =\left(\frac{2(\vec{r}_{12'}
\vec{r}_{22'} )}{\vec{r}_{12'}^{\;2}\vec{r}_{22'}^{\;2}}
-\frac{1}{\vec{r}_{12'}^{\;2}}\right)
\ln\left(\frac{\vec{r}_{12'}^{\;2}}{\vec{r}_{12}^{\;2}}\right)
\ln\left(\frac{\vec{r}_{22'}
^{\;2}}{\vec{r}_{12}^{\;2}}\right)-\frac{1}{\vec{r}_{22'}^{\;2}}
\ln^2\left(\frac
{\vec{r}_{12'}^{\;2}}{\vec{r}_{12}^{\;2}}\right) \label{J coordinate}
\end{equation}
and
\[
I(\vec{r}_{1}, \vec{r}_{2},\vec{r}_{1}^{\;\prime},\vec{r}_{2}^{\;\prime})
=\frac{1}{\vec{r}_{1'2'}^{\;2}}\left(\frac{(\vec{r}_{11'}\vec{r}_{22'} )}
{\vec{r}_{11'}^{\;2}\vec{r}_{22'}^{\;2}}\ln\left(\frac{\vec{r}_{21'}^{\;2}
\vec{r}_{1'2'}^{\;2}}
{\vec{r}_{12}^{\;2}\vec{r}_{12'}^{\;2}}\right)+\frac{(\vec{r}_{11'}
\vec{r}_{12'} )}{\vec{r}_{11'}^{\;2}\vec{r}_{12'}^{\;2}}
\ln\left(\frac{\vec{r}_{12'}^{\;4}\vec{r}_{12}^{\;2}}
{\vec{r}_{11'}^{\;2}\vec{r}_{22'}^{\;4}}\right)\right.
\]
\begin{equation}
\left. +\frac{(\vec{r}_{22'}\vec{r}_{21'})}{\vec{r}_{22'}^{\;2}
\vec{r}_{21'}^{\;2}}
\ln\left(\frac{\vec{r}_{12}^{\;2}}{\vec{r}_{21'}^{\;2}}\right)
+\frac{(\vec{r}_{12'}
\vec{r}_{21'})}{\vec{r}_{12'}^{\;2}\vec{r}_{21'}^{\;2}}
\ln\left(\frac{\vec{r}_{11'}^{\;2}
\vec{r}_{22'}^{\;2}}{\vec{r}_{12}^{\;2}\vec{r}_{1'2'}^{\;2}}\right)\right)~.
\label{I coordinate}
\end{equation}
Note that the property of turning into zero at $\vec{r}_{12}=0$ is conserved 
after integration. Thus, the M\"{o}bius form of the commutator given
by Eqs.~(\ref{mebius commutator}),~(\ref{F coordinate}),~(\ref{J coordinate})
and~(\ref{I coordinate}) is rather simple and does not contain special
functions.
Having this form one can find the complete commutator in the momentum space
$\langle\vec{q}_{1},\vec{q}_{2}|\alpha_s\left[\hat{\mathcal{K}}^{(B)},\hat
{U}\right]|\vec{q}_{1}^{\,\,\prime},\vec{q}_{2}^{\,\,\prime}\rangle$
according to the prescriptions of Ref.~\cite{Fadin:2011jg}. We write it in the
form
\[
\langle\vec{q}_{1},\vec{q}_{2}|\alpha_s\left[\hat{\mathcal{K}}^{(B)},\hat
{U}\right]|\vec{q}_{1}^{\,\,\prime}%
,\vec{q}_{2}^{\,\,\prime}\rangle=\delta(\vec{q}_{11^{\prime}}+\vec
{q}_{22^{\prime}}) \frac{\alpha^2_s N^2_c}{8\pi^{3}} \Biggl[\frac{\beta_0}
{4N_c} \ln\left(\frac{\vec{q}_1^{\;2}}{\vec{q}_{1}^{\;\prime\,2}}
\right) R(\vec{q}_1, \vec{q}_2; \vec{k})
\]
\begin{equation}
+ F(\vec{q}_{2}, \vec{q}_{2}; \vec{k})+J(\vec{q}_{2}, \vec{q}_{2}; \vec{k})
+I(\vec{q}_{2}, \vec{q}_{2}; \vec{k})+\vec{q}_1\leftrightarrow -\vec{q}_2
\biggr]~.
\label{restored commutator}
\end{equation}
Here $R(\vec{q}_1, \vec{q}_2; \vec{k})$ is given by 
Eq.~(\ref{R  momentum}) and
\[
F(\vec{q}_{1}, \vec{q}_{2}; \vec{k})=\frac{1}{\pi}<\int\frac{d\vec{r}_{11'}}
{2\pi}
\frac{d\vec{r}_{22'}}{2\pi}{d\vec{r}_{1'2'}}e^{-i(\vec{q}_1
\vec{r}_{11'}+\vec{q}_2\vec{r}_{22'}+\vec{k}\vec{r}_{1'2'})}F(\vec{r}_{1},
\vec{r}_{2},\vec{r}_{1}^{\;\prime}, \vec{r}_{2}^{\;\prime})>
\]
\[
=\left(\frac{1}{\vec{q}_{1}^{\;2}}\ln\left(\frac{\vec{q}_1^{\;2}}
{\vec{q}_1^{\;\prime\,2}}\right)
\ln\left(\frac{\vec{q}_2^{\;2}}{\vec{q}^{\;2}}\right)
+\frac{1}{2 \vec{k}^{\;2}}\ln\left(\frac{\vec{q}_1^{\;2}}
{\vec{q}_1^{\;\prime\,2}}\right)
\ln\left(\frac{\vec{q}_2^{\;2}}{\vec{q}_2^{\;\prime\,2}}\right)
-\frac{(\vec{q}_{1}\vec{k})}{\vec{q}_{1}^{\;2}
\vec{k}^{\;2}}\right.
\]
\[
\left.
\times
\left(\ln\left(\frac{\vec{q}_2^{\;\prime\,2}}
{\vec{q}_2^{\;2}}\right)\ln\left(\frac{\vec{q}_1^{\;\prime\,2}}
{\vec{k}^{\;2}}\right)+\ln\left(\frac{\vec{q}^{\;2}}
{\vec{q}_2^{\;2}}\right)\ln\left(\frac{\vec{q}_2^{\;2}\vec{q}_1^{\;\prime\,2}}
{\vec{q}_1^{\;2}\vec{q}_2^{\;\prime\,2}}\right)\right)
+\vec{q}_1\leftrightarrow -\vec{q}_2\right)+\frac{(\vec{q}_{1}\vec{q}_{2})}
{\vec{q}_{1}^{\;2}
\vec{q}_{2}^{\;2}}
\]
\[
\times\left(\ln\left(\frac{\vec{q}^{\;2}}{\vec{q}_1^{\;2}}\right)
\ln\left(\frac{\vec{q}_1^{\;\prime\,2}\vec{k}^{\;2}}
{\vec{q}_1^{\;2}\vec{q}_2^{\;2}}\right)
+\ln\left(\frac{\vec{q}_2^{\;\prime2}}{\vec{k}^{\;2}}\right)
\ln\left(\frac{\vec{q}^{\;2}}
{\vec{q}_1^{\;\prime2}}\right)\right)
-2
\frac{[\vec{q}_{1}\times\vec{q}_2]}{\vec{q}_{1}^{\;2}\vec{q}_2^{\;2}}
[\vec{q}_{1}\times\vec{k}]
I_{\vec{k}, \vec{q}_{1}^{\;\prime}}
\]
\begin{equation}
-2\left(\frac{[\vec{q}_{1}\times\vec{k}]}
{\vec{q}_{1}^{\;2}\vec{k}^{\;2}}+\frac{[\vec{q}_{2}\times\vec{k}]}
{\vec{q}_{2}^{\;2}\vec{k}^{\;2}}+
\frac{[\vec{q}_{1}\times\vec{q}_2]}{\vec{q}_{1}^{\;2}\vec{q}_2^{\;2}}\right)
\left([\vec{q}_{1}\times\vec{q}_2]I_{\vec{q}_{1}, \vec{q}_{2}}-
[\vec{q}_{1}^{\;\prime}\times\vec{q}_2^{\;\prime}]I_{\vec{q}_{1}^{\;\prime},
\vec{q}_{2}^{\;\prime}}\right)\;,  \label{F momentum}
\end{equation}
\begin{equation}
J(\vec{q}_{2}, \vec{q}_{2}; \vec{k})=\frac{1}{\pi}<\int\frac{d\vec{r}_{11'}}
{2\pi}\frac{d\vec{r}_{22'}}{2\pi}{d\vec{r}_{1'2'}}e^{-i(\vec{q}_1
\vec{r}_{11'}+\vec{q}_2\vec{r}_{22'}+\vec{k}\vec{r}_{1'2'})}
\delta(\vec{r}_{11^{\prime}})J(\vec{r}_{1},\vec{r}_{2},\vec{r}_{2}^{\;\prime})>
\]
\[
\left(\frac{1}{\vec{q}_{2}^{\;2}}+2\frac{(\vec{q}_{2}\vec{k})}
{\vec{q}_{2}^{\;2}
\vec{k}^{\;2}}\right)\ln\left(\frac{\vec{q}_2^{\;2}}
{\vec{q}_2^{\;\prime2}}\right)
\ln\left(\frac{\vec{q}_2^{\;\prime\,2}}{\vec{k}^{\;2}}\right)
-\frac{1}{\vec{k}^{\;2}}
\ln^2\left(\frac{\vec{q}_2^{\;2}}
{\vec{q}_2^{\;\prime2}}\right)~, \label{J momentum}
\end{equation}
\[
I(\vec{q}_{1}, \vec{q}_{2}; \vec{k})=\frac{1}{\pi}<\int\frac{d\vec{r}_{11'}}
{2\pi}\frac{d\vec{r}_{22'}}{2\pi}{d\vec{r}_{1'2'}}e^{-i(\vec{q}_1
\vec{r}_{11'}+\vec{q}_2\vec{r}_{22'}+\vec{k}\vec{r}_{1'2'})}F(\vec{r}_{1},
\vec{r}_{2},\vec{r}_{1}^{\;\prime}, \vec{r}_{2}^{\;\prime})>
\]
\[
=
\frac{1}{2\vec{q}_{1}^{\;2}}\ln\left(\frac{\vec{q}_1^{\;\prime\,2}}
{\vec{q}^{\;2}}\right)\ln\left(\frac{\vec{q}_2^{\;\prime\,2}}
{\vec{q}^{\;2}}\right)+\frac{1}{2\vec{q}_{2}^{\;2}}
\left(\ln\left(\frac{\vec{q}_1^{\;\prime\,2}}
{\vec{q}^{\;2}}\right)\ln\left(\frac{\vec{q}_2^{\;\prime\,2}}
{\vec{q}^{\;2}}\right)\right.
\]
\[
\left.-2\ln\left(\frac{\vec{q}_1^{\;\prime\,2}}
{\vec{q}_1^{\;2}}\right)\ln\left(\frac{\vec{k}^{\;2}}
{\vec{q}_1^{\;2}}\right)\right)
 -\frac{(\vec{q}_{1}\vec{q}_{2})}{\vec{q}_{1}^{\;2}
\vec{q}_{2}^{\;2}}\left(\ln\left(\frac{\vec{q}^{\;2}}{\vec{q}_1^{\;2}}\right)
\ln\left(\frac{\vec{q}_1^{\;\prime\,2}\vec{k}^{\;2}}
{\vec{q}_1^{\;2}\vec{q}_2^{\;2}}\right)
+\ln\left(\frac{\vec{q}_2^{\;\prime2}}{\vec{k}^{\;2}}\right)
\ln\left(\frac{\vec{q}^{\;2}}
{\vec{q}_1^{\;\prime2}}\right)\right)
\]
\begin{equation}
+2\frac{[\vec{q}_{1}\times\vec{q}_2]}{\vec{q}_{1}^{\;2}\vec{q}_2^{\;2}}
[\vec{q}_{1}\times\vec{k}]
I_{\vec{k}, \vec{q}_{1}^{\;\prime}}-2\frac{[\vec{q}_{1}
\times\vec{q}_2]}{\vec{q}_{1}^{\;2}
\vec{q}_2^{\;2}}[\vec{q}_{1}^{\;\prime}\times\vec{q}_2^{\;\prime}]I_{\vec{q}_{1}^{\;\prime}, \vec{q}_{2}^{\;\prime}}~.  \label{I momentum}
\end{equation}
In these equalities the symbols $<.....>$ mean adding to the direct 
Fourier transform terms
that depend only on $\vec{q}_{1}$ and $\vec{q}_{2}$ (and do not depend on
$\vec k$) and terms that are antisymmetric with respect to the substitution
$\vec{q}_1\leftrightarrow -\vec{q}_2$. These terms are fixed by the
requirement of the gauge invariance and the symmetry of the kernel,
according to Ref.~\cite{Fadin:2011jg}.

Equalities~(\ref{F momentum})--(\ref{I momentum}) can be derived using
formulas given in the Appendices of Ref.~\cite{Fadin:2007de} and of the
present paper. The substitution of these equalities
in Eq.~(\ref{restored commutator}) gives the same result 
as Eq.~(\ref{final commutator}).

\section{Conclusion}
\label{sec:conclusion}

The simplicity of the M\"{o}bius form of the quasi-conformal NLO BFKL
kernel suggested to use  just this form  for finding the kernel in the
momentum space.  The way to do that was not evident,  and even the
possibility to do it seemed  doubtful, because the M\"{o}bius form is defined
on a special class of functions in the coordinate space. However, it was
shown~\cite{Fadin:2011jg} that such possibility exists due to the gauge
invariance of the kernel and the way to obtain the kernel in the momentum
space from its M\"{o}bius form was elaborated. But technically obtaining
it turned out to be not easy. 

In this paper we found in the momentum space the difference between the
standard BFKL kernel, defined according to the prescriptions given in
Ref.~\cite{Fadin:1998fv} and calculated in Ref.~\cite{FF05},  and the
quasi-conformal BFKL kernel. This difference turned out to be rather simple.
The most natural conclusion is that the simplicity of the M\"{o}bius form
of the quasi-conformal kernel is caused mainly by using the impact parameter
space. The other possibility is that the quasi-conformal kernel can be 
written in simple form also in the transverse momentum space. If this is
true, the standard kernel of Ref.~\cite{FF05} could result itself in a much 
simpler form. We plan to check this possibility using
both the representation of Ref.~\cite{FF05} and the representation in terms
of integrals in the transverse momentum space of Ref.~\cite{Fadin:2006zz}.

\vspace{0.5cm} {\textbf{{\Large Acknowledgments}}}

\vspace{0.5cm} V.S.F. thanks the Dipartimento di Fisica
dell'Universit\`{a} della Calabria and the Istituto Nazionale di
Fisica Nucleare (INFN), Gruppo Collegato di Cosenza, for warm hospitality
while part of this work was done and for financial support.

\newpage
\setcounter{equation}{0}
\def\theequation{A.\arabic{equation}}
\section*{Appendix}

The two-dimensional integrals of Section~\ref{sec:direct} were calculated
choosing appropriate integration vectors and  performing firstly the
integration over azimuthal angles.  It is convenient to make this integration
using ``helical'' vector components ``$\pm$'' instead of the
Cartesian ones ``$x,y$'', $a^{\pm}=a_x\pm i a_y$.
Denoting the integration vector as $\vec{l}$, we have
$l^\pm = le^{\pm i \phi}$, where $\phi$ is its azimuthal angle and $l$ is
its modulus. The integration over $\phi$ can be performed using
the representation $2(\vec{a}-\vec{l})(\vec{b}-\vec{l})=(a^+-l^+)(b^--l^-)
+(a^--l^-)(b^+-l^+)$
and the expansion of the integrands in positive or negative powers of
$l^\pm$ at various values of $l$. Thus one can easily obtain
\begin{equation}
\int_{-\pi}^{\pi}\frac{d\phi}{2\pi}\;\ln(\vec{a}-\vec{l})^2
=\theta(\vec{a}^{\;2}-\vec{l}^{\;2})\ln\vec{a}^{\;2}
+\theta(\vec{a}^{\;2}-\vec{l}^{\;2})\ln\vec{l}^{\;2}~,\label{int phi ln}
\end{equation}
\begin{equation}
\int_{-\pi}^{\pi}\frac{d\phi}{2\pi}\;\frac{1}{a^\pm-l^\pm}=
\frac{\theta(\vec{a}^{\;2}-\vec{l}^{\;2})}{a^\pm}~,\;\;\label{int phi pm}
\end{equation}
\begin{equation}
\int_{-\pi}^{\pi}\frac{d\phi}{2\pi}\;\frac{1}{(a^\pm -l^\pm)
(b^\mp -l^\mp)}=
\frac{\theta(\vec{a}^{\;2}-\vec{l}^{\;2})}{a^\pm b^\mp-\vec{l}^{\;2}}+
\frac{\theta(\vec{l}^{\;2}-\vec{b}^{\;2})}{\vec{l}^{\;2}-a^\pm b^\mp}~.
\label{int phi pm mp}
\end{equation}
In particular, one has from Eq.~(\ref{int phi pm mp})
\[\int_{-\pi}^{\pi}\frac{d\phi}{2\pi}\;\frac{2(\vec{l}
(\vec{a}-\vec{l}))}{\vec{l}^{\;2}
(\vec{a}-\vec{l})^{2}}=\theta(\vec{l}^{\;2}-\vec{a}^{\;2})\frac{-2}
{\vec{l}^{\;2}}~,\;\;
\]
\begin{equation}
\int_{-\pi}^{\pi}\frac{d\phi}{2\pi}\;\frac{2((\vec{a}-\vec{l}))
(\vec{b}-\vec{l}))}{(\vec{a}-\vec{l})^{2}(\vec{b}-\vec{l})^{2}}
=\left(\theta(\vec{a}^{\;2}-\vec{l}^{\;2})-\theta(\vec{b}^{\;2}-\vec{l}^{\;2})
\right)\left(\frac{1}{a^+b^--\vec{l}^{\;2}}
+\frac{1}{a^-b^+-\vec{l}^{\;2}}\right)~. \label{int phi a-l b-l}
\end{equation}
The result~(\ref{int R u}) follows from Eqs.~(\ref{int phi ln})
and~(\ref{int phi a-l b-l}) with the subsequent elementary integration over
$l$. Since the integral consists of several terms, which are not ultraviolet
convergent when taken separately, it is convenient to calculate them
introducing an ultraviolet cut-off $\Lambda$.
Using Eq.~(\ref{int phi a-l b-l}), one can also easily obtain
\begin{equation}
\int\frac{d{\vec{l}}}{\pi}\;\left(\frac{(\vec{l}(\vec{a}-\vec{l}))}
{\vec{l}^{\;2}
(\vec{a}-\vec{l})^{2}}-\frac{(\vec{l}(\vec{b}-\vec{l}))}{\vec{l}^{\;2}
(\vec{b}-\vec{l})^{2}}\right)=\ln\left(\frac{\vec{b}^{\;2}}{\vec{a}^{\;2}}
\right)~, \label{int v}
\end{equation}
that gives the result~(\ref{virtual contribution}).

Though we use the ultraviolet cut-off $\Lambda$ (which is supposed tending to
infinity) for separate integrals, it is possible to shift the integration
vectors in them, since these integrals have only logarithmic divergence.
Therefore, with an appropriate choice of $\vec{l}$, in all integrals of
Section~\ref{sec:direct} the integration over $\phi$ can be performed
using Eqs.~(\ref{int phi pm}) and~(\ref{int phi pm mp}). But sometimes it
is more convenient to use Eq.~(\ref{int phi ln}) as, for example, in the 
integral
\begin{equation}
\int\frac{d{\vec{l}}}{\pi}\;\theta(\Lambda^2-\vec{l}^{\;2})
\frac{1}{(\vec{a}-\vec{l})^2}\ln\left(\frac{\vec{l}^{\;2}}{\vec{a}^{\;2}}
\right) =\int\frac{d{\vec{l}}}{\pi}\;\theta(\Lambda^2-\vec{l}^{\;2})
\frac{1}{\vec{l}^{\;2}}\ln\frac{(\vec{a}-\vec{l})^2}{\vec{a}^{\;2}} =
\frac{1}{2}\ln^2\left(\frac{\Lambda^2}{\vec{a}^{\;2}}\right)~. \label{int s1l}
\end{equation}
Using Eq.~(\ref{int phi pm mp}), we  obtain:
\[
\int\frac{d \vec l}{\pi}\frac{1}{(\vec a- \vec 1)^+}
\frac{1}{(\vec b- \vec 1)^-}\ln\left(\frac{\vec l^{\;2}}{\mu^2}\right)
\theta(\Lambda^2-\vec l^{\;2})=
\frac12\ln\left(\frac{\Lambda^2}{(\vec a -\vec b)^{2}}\right)
\ln\left(\frac{\Lambda^2(\vec a -\vec b)^{2}}{\mu^4}\right)
\]
\begin{equation}
+\frac12\ln\left(\frac{(\vec a -\vec b)^{2}}{\vec b^{\;2}}\right)
\ln\left(\frac{(\vec a -\vec b)^{2}}{\vec a^{\;2}}\right)
+\frac{a^+b^--a^-b^+}{2}I_{\vec a,-\vec b }~, \label{int master}
\end{equation}
where $I_{\vec a,\vec b }$ is defined in Eq.~(\ref{I p q 1})
(see also Eqs.~(\ref{I symmetric}) and~(\ref{integral I})). In fact, all 
integrals of Section~\ref{sec:direct} can be calculated using this one. In 
particular, the integral~(\ref{int s1l}) can be obtained from 
the integral~(\ref{int master}) as the limit $\vec b \rightarrow \vec a$ at 
$\mu^2=\vec a^{\;2}$. The integrals~(\ref{int phi a-l b-l}) and~(\ref{int v}) 
also can be found using the part of the integral~(\ref{int master}) 
proportional to $\ln \mu^2$. We find also
\[
\int\frac{d\vec l}{\pi}\frac{2}{(\vec a-\vec l)^2}\frac{(\vec b-\vec l)}
{(\vec b-\vec l)^2}
\ln\left(\frac{\vec l^{\;2}}{\vec a^{\;2}}\right) =\frac{(\vec a-\vec b)}
{(\vec a-\vec b)^{2}}\ln\left(\frac{\vec a^{\;2}}{\vec b^{\;2}}\right)
\ln\left(\frac{(\vec a -\vec b)^{2}}{\vec a^{\;2}}\right)
\]
\begin{equation}
+2\frac{[(\vec a- \vec b)\times[\vec a \times\vec b]]}{(\vec a-\vec b)^2}
I_{\vec a, -\vec b}~,
\label{int vl2}
\end{equation}
\[
\int\frac{d\vec l}{\pi}\frac{(\vec{c}-\vec l)}{(\vec{c}-\vec l)^2}
\frac{2((\vec a-\vec l)(\vec b-\vec l))}{(\vec a -\vec l)^2(\vec b-\vec l)^2}
\ln\left(\frac{\vec l^{\;2}}{\mu^{2}}\right) =\frac{(\vec c -\vec b)}
{(\vec c -\vec b)^2}\left[\ln\left(\frac{\vec a^{\;2}}{\mu^{2}}\right)
\ln\left(\frac{(\vec c -\vec a)^{2}}{(\vec b -\vec a)^{2}}\right)\right.
\]
\[
\left.+\frac12\ln\left(\frac{c^{\;2}}{\vec a^{\;2}}\right)
\ln\left(\frac{(\vec a -\vec c)^{2}}{\vec b^{\;2}}\right)
-\frac12\ln\left(\frac{b^{\;2}}{\vec a^{\;2}}\right)
\ln\left(\frac{(\vec a -\vec b)^{2}}{\vec c^{\;2}}\right)
\right] +\frac{(\vec c -\vec a)}{(\vec c -\vec a)^2}
\left[\ln\left(\frac{\vec b^{\;2}}{\mu^{2}}\right)\right.
\]
\[
\left.
\times \ln\left(\frac{(\vec c -\vec b)^{2}}{(\vec a -\vec b)^{2}}\right)
+\frac12\ln\left(\frac{c^{\;2}}{\vec b^{\;2}}\right)
\ln\left(\frac{(\vec b -\vec c)^{2}}{\vec a^{\;2}}\right)
-\frac12\ln\left(\frac{a^{\;2}}{\vec b^{\;2}}\right)
\ln\left(\frac{(\vec b -\vec a)^{2}}{\vec c^{\;2}}\right)
\right]
\]
\[
+\left(\frac{[(\vec c -\vec b)\times[\vec a\times\vec b]]}
{(\vec c -\vec b)^2}+\frac{[(\vec c -\vec a)\times[\vec b\times\vec a]]}
{(\vec c -\vec a)^2}\right)I_{\vec a,-\vec b }
+\frac{[(\vec c -\vec b)\times[\vec c\times\vec a]]}
{(\vec c -\vec b)^2}I_{\vec c,-\vec a }
\]
\begin{equation}
+\frac{[(\vec c -\vec a)\times[\vec c\times\vec b]]}
{(\vec c -\vec a)^2}I_{\vec c,-\vec b }\label{int vl3}~.
\end{equation}
The result~(\ref{F1 final}) for $F_1(\vec{q}_1,\vec{q}_2; \vec{k})$ was
obtained using Eqs.~(\ref{int R u}),~(\ref{int vl2}) and~(\ref{int vl3}) with
its particular cases, such as
\[
\int\frac{d\vec l}{\pi}\frac{(\vec{a}-\vec l)}{(\vec{a}-\vec l)^2}
\frac{2(\vec l(\vec b-\vec l))}{\vec l^{\;2}(\vec b-\vec l)^2}
\ln\left(\frac{\vec l^{\;2}}{\mu^{2}}\right) =-\frac12
\frac{(\vec a -\vec b)}{(\vec a -\vec b)^2}\ln\left(\frac{\vec a^{\;2}}
{\vec b^{\;2}}\right)\ln\left(\frac{\vec a^{\;2}\vec b^{\;2}}{\mu^{4}}\right)
\]
\begin{equation}
 -\frac12\frac{\vec a}{\vec a^{\;2}}\ln\left(\frac{\vec a^{\;2}\vec b^{\;2}}
{\mu^{4}}\right) \ln\left(\frac{(\vec a -\vec b)^{2}}{\vec b^{\;2}}\right)
-\frac{[\vec a\times[\vec a\times\vec b]]}{\vec a^{\;2}}I_{\vec a,-\vec b }~.
\label{int vl30}
\end{equation}
To obtain  $F_2(\vec{q}_1,\vec{q}_2; \vec{k})$, Eq.~(\ref{F2 final}), we used
\[
\int\frac{d\vec l}{\pi}\theta(\Lambda^2-\vec l^{\;2})\;\frac{2((\vec a-\vec l)
(\vec b-\vec l))}{(\vec a-\vec l)^{2}(\vec b-\vec l)^2}
\ln\left(\frac{\vec l^{\;2}}{\mu^{2}}\right)
=\ln\left(\frac{\Lambda^2(\vec a -\vec b)^2}{\mu^4}\right)
\ln\left(\frac{\Lambda^2}{(\vec a -\vec b)^2}\right)
\]
\begin{equation}
+\ln\left(\frac{\vec a^{\;2}}{(\vec a -\vec b)^2}\right)
\ln\left(\frac{\vec b^{\;2}}{(\vec a -\vec b)^2}\right)~,
\label{int s2l}
\end{equation}
\[
\int\frac{d\vec l}{\pi}\;\frac{1}{\vec l^{\;2}}\frac{2((\vec a-\vec l)
(\vec b-\vec l))}{(\vec a-\vec l)^2(\vec b-\vec l)^2}
\ln\left(\frac{(\vec c-\vec l)^2}{\vec c^{\;2}}\right) =\frac{1}{\vec a^{\;2}
\vec b^{\;2}}\left[(\vec a\vec b)\left(\ln\left(\frac{(\vec c -\vec a)^{2}}
{\vec c^{\;2}}\right)\ln\left(\frac{(\vec c -\vec b)^{2}}{\vec c^{\;2}}\right)
\right.\right.
\]
\[
\left.\left.-\ln\left(\frac{(\vec c -\vec a)^{2}}{\vec c^{\;2}}\right)
\ln\left(\frac{(\vec a -\vec b)^{2}}{\vec a^{\;2}}\right)
-\ln\left(\frac{(\vec c -\vec b)^{2}}{\vec c^{\;2}}\right)
\ln\left(\frac{(\vec a -\vec b)^{2}}{\vec b^{\;2}}\right)\right)\right.
\]
\[
\left.
+2([\vec a\times\vec b][\vec a\times\vec c]) I_{\vec a, -\vec c}
+2([\vec a\times\vec b][\vec c\times\vec b])I_{\vec b, -\vec c}
\right.
\]
\begin{equation}
\left.+2([\vec a\times\vec b][(\vec a-\vec c)\times(\vec b-\vec c)])
I_{\vec a -\vec c, \vec c-\vec b }\right]~,
\label{int s3l}
\end{equation}
in particular,
\[
\int\frac{d\vec l}{\pi}\frac{1}{(\vec a-\vec l)^{2}}
\frac{2(\vec l(\vec b-\vec l))}{\vec l^{\;2}(\vec b-\vec l)^2}
\ln\left(\frac{\vec l^{\;2}}{\vec a^{\;2}}\right) =\frac{1}{\vec a^{\;2}
(\vec a-\vec b)^{2}}\left[(\vec a(\vec a-\vec b))\ln\left(\frac{\vec a^{\;2}}
{\vec a^{\;2}}\right)\ln\left(\frac{(\vec a -\vec b)^{2}}{\vec b^{\;2}}\right)
\right.
\]
\begin{equation}
\left.-2[\vec a \times\vec b]^2I_{\vec a -\vec b, -\vec a }\right]~,
\label{int s3l0}
\end{equation}
and Eq.~(\ref{int vl3}) with its particular cases~(\ref{int vl30}) and
\[
\int\frac{d\vec l}{\pi}\frac{\vec l}{\vec l^{\;2}}\frac{2((\vec a-\vec l)
(\vec b-\vec l))}{(\vec a -\vec l)^2(\vec b-\vec l)^2}
\ln\left(\frac{\vec l^{\;2}}{\mu^{2}}\right) =\frac{\vec b}{\vec b^{\;2}}
\left[\ln\left(\frac{\vec a^{\;2}}{\mu^{2}}\right)\ln\left(\frac{\vec a^{\;2}}
{(\vec b -\vec a)^{2}}\right)\right.
\]
\[
\left.+\frac12\ln\left(\frac{(\vec b -\vec a)^{2}}{\vec a^{\;2}}\right)
\ln\left(\frac{\vec a^{\;2}}{\vec b^{\;2}}\right)\right] +\frac{\vec a}
{\vec a^{\;2}}\left[\ln\left(\frac{\vec b^{\;2}}{\mu^{2}}\right)
\ln\left(\frac{\vec b^{\;2}}{(\vec a -\vec b)^{2}}\right)\right.
\]
\begin{equation}
\left.
 +\frac12\ln\left(\frac{(\vec a -\vec b)^{2}}{\vec b^{\;2}}\right)
\ln\left(\frac{\vec b^{\;2}}{\vec a^{\;2}}\right)
\right]
+\left(\frac{[\vec b\times[\vec a\times\vec b]]}{\vec b^{\;2}}
+\frac{[\vec a\times[\vec b\times\vec a]]}{\vec a^{\;2}}\right)
I_{\vec a,-\vec b }~.
\label{int v3l0}
\end{equation}
The result~(\ref{F3 final}) for $F_3(\vec{q}_1,\vec{q}_2; \vec{k})$ can be
obtained using Eqs.~(\ref{int vl2}),~(\ref{int s3l0}),~(\ref{int v3l0}),
\begin{equation}
\int\frac{d\vec l}{\pi}\frac{1}{\vec l^{\;2}(\vec a-\vec l)^2}
\ln\left(\frac{(\vec b-\vec l)^{2}(\vec a-\vec b-\vec l)^{2}}{\vec b^{\;2}
(\vec a-\vec b)^{2}}\right) =\frac{1}{\vec a^{\;2}}\ln^2\left(\frac{(\vec a
-\vec b)^{2}}{\vec b^{\;2}}\right)\label{int s2ll}
\end{equation}
and
\[
\int\frac{d\vec l}{\pi}\;\frac{\theta(\Lambda^2-\vec l^{\;2})}
{(\vec c-\vec l)^2}\left(\frac{2((\vec a-\vec l)(\vec b-\vec l))}
{(\vec a-\vec l)^2(\vec b-\vec l)^2}-2\frac{2((\vec a-\vec c)
(\vec b-\vec c))}{(\vec a-\vec c)^2(\vec b-\vec c)^2}\right)
\ln\left(\frac{\vec l^{\;2}}{\mu^{2}}\right)
\]
\[
=\frac{((\vec a-\vec c)(\vec b-\vec c))}{(\vec a-\vec c)^2(\vec b-\vec c)^2}
\left[\ln\left(\frac{\Lambda^2}{(\vec a -\vec b)^{2}}\right)
\ln\left(\frac{\Lambda^2(\vec a -\vec b)^{2}}{\mu^4}\right)
+\ln\left(\frac{(\vec a -\vec b)^{2}}{\vec a^{\;2}}\right)
\ln\left(\frac{(\vec a -\vec b)^{2}}{\vec b^{\;2}}\right)\right.
\]
\[
\left.
-\ln\left(\frac{\Lambda^2}{(\vec a -\vec c)^{2}}\right)
\ln\left(\frac{\Lambda^2(\vec a -\vec c)^{2}}{\mu^4}\right)
- \ln\left(\frac{(\vec a -\vec c)^{2}}{\vec a^{\;2}}\right)
\ln\left(\frac{(\vec a -\vec c)^{2}}{\vec c^{\;2}}\right)
\right.
\]
\[
\left.
-\ln\left(\frac{\Lambda^2}{(\vec c -\vec b)^{2}}\right)
\ln\left(\frac{\Lambda^2(\vec c -\vec b)^{2}}{\mu^4}\right)
-\ln\left(\frac{(\vec c -\vec b)^{2}}{\vec c^{\;2}}\right)
\ln\left(\frac{(\vec c -\vec b)^{2}}{\vec b^{\;2}}\right)\right]
\]
\begin{equation}
+2\left(\frac{[(\vec a-\vec c)\times(\vec b-\vec c)]}
{(\vec a-\vec c)^2(\vec b-\vec c)^2}\left([\vec a\times\vec b]
I_{\vec a, -\vec b}-[\vec a\times\vec c]
I_{\vec a, -\vec c} -[\vec c\times\vec b] I_{\vec c, -\vec b}   \right)\right)\;.
\label{int s3ld}
\end{equation}
Let us present also the integral
\[
\int\frac{d\vec l}{\pi}\frac{2}{(\vec a-\vec l)^2}\frac{l_i(b-\vec l)_j}
{\vec l^{\;2}(\vec b-\vec l)^2}
\ln\left(\frac{\vec l^{\;2}}{\vec a^{\;2}}\right)
=\frac{b_i(a-b)_j+(a-b)_i b_j-\delta_{ij}(\vec b(\vec a-\vec b))}
{2\vec b^{\;2}(\vec a-\vec b)^{2}}\ln\left(\frac{\vec a^{\;2}}
{\vec b^{\;2}}\right)
\]
\[
\times \ln\left(\frac{(\vec a -\vec b)^{2}}{\vec a^{\;2}}\right)
+\frac{b_ia_j-a_i b_j+\delta_{ij}(\vec a(\vec a-\vec b))}{2\vec a^{\;2}
(\vec a-\vec b)^{2}}\ln\left(\frac{\vec a^{\;2}}{\vec b^{\;2}}\right)
\ln\left(\frac{(\vec a -\vec b)^{2}}{\vec b^{\;2}}\right)
+\frac{I_{\vec a, -\vec b}}{(\vec a-\vec b)^2}
\]
\[
\times\left(\frac{1}{\vec b^{\;2}}
\left([\vec b\times[\vec a\times \vec b]]_i(\vec a- \vec b)_j
+(\vec a- \vec b)_i[\vec b\times[\vec a\times \vec b]]_j
-\delta_{ij}([\vec b\times[\vec a\times \vec b]](\vec a- \vec b))
\right)\right.
\]
\begin{equation}
\left.+\frac{1}{\vec a^{\;2}}\left( (\vec a- \vec b)_i[\vec a
\times[\vec a\times \vec b]]_j-[\vec a\times[\vec a\times \vec b]]_i(\vec a
- \vec b)_j -\delta_{ij}([\vec a\times[\vec a\times \vec b]](\vec a- \vec b))
\right)\right)
\label{int tl3}
\end{equation}
which is more general than the integral~(\ref{int vl30}) and can appear  
in decompositions of the integrands for $F_i$ different from ours, and the 
integrals
\[
\int\frac{d\vec l}{\pi}\frac{2}{\vec l^{\;2}}\left(\frac{(\vec a-\vec l)}{(\vec a-\vec l)^2}
\left(\frac{(\vec b-\vec l)}{(\vec b-\vec l)^2}-\frac{\vec b}{\vec b^{\;2}}
\right)\right)
\ln\left(\frac{\vec l^{\;2}}{\vec q^{\;2}}\right) =\frac{(\vec a\vec b)}
{\vec a^{\;2}\vec b^{\;2}}\ln\left(\frac{\vec a^{\;2}\vec b^{\;2}}
{\vec q^{\;4}}\right)\ln\left(\frac{\vec b^{\;2}}{(\vec a -\vec b)^{2}}\right)
\]
\begin{equation}
+\frac{2[\vec a\times\vec b]^{2}}{\vec a^{\;2}\vec b^{\;2}}
I_{\vec a, -\vec b}~,
\label{int sl3d}
\end{equation}
\[
\int\frac{d\vec l}{\pi}\frac{2}{\vec l^{\;2}}\left(\frac{(\vec b-\vec l)}
{(\vec b-\vec l)^2} -\frac{\vec b}{\vec b^{\;2}} \right)
\ln\left(\frac{(\vec a-\vec l)^2}{\vec l^{\;2}}\right)
=\frac{\vec b}{\vec b^{\;2}}\ln\left(\frac{\vec a^{\;2}}{\vec b^{\;2}}\right)
\ln\left(\frac{\vec b^{\;2}}{(\vec a -\vec b)^{2}}\right)
\]
\begin{equation}
+2\frac{[\vec b\times[\vec a \times\vec b]]}{\vec b^{\;2}}
I_{a,\vec b-\vec a }~,\label{int vl2d}
\end{equation}
which also can be useful.

\end{document}